\documentclass[copyright,creativecommons]{eptcs}
\providecommand{\event}{SNR 2017} 
\usepackage{underscore}                    

\usepackage{amsmath}
\usepackage{amssymb}
\DeclareMathOperator{\range}{range}
\newcommand{\R}{\mathbb{R}} 
\newcommand{\B}{\mathbb{B}}  
\newcommand{\Epsilon}{\mathbb{E}} 
\newcommand{\X}{\mathbb{X}} 
\newcommand{\x}{\chi}             
\usepackage[pdftex]{graphicx}
\DeclareGraphicsExtensions{.pdf,.jpg,.png}
\usepackage{listings, color}
\usepackage{parcolumns}
\usepackage{subfig}
\newtheorem{definition}{\bf Definition}[section]

\usepackage{todonotes}
\usepackage{soul}
\usepackage{xspace}
\usepackage{siunitx}

\definecolor{lightblue}{rgb}{.9,.95,1}

\newcommand{\uE}{\ensuremath{\dot{E}}\xspace\xspace}
\newcommand{\uS}{\ensuremath{\dot{S}}\xspace}

\newcommand{\us}{\ensuremath{\dot{s}}\xspace}

\newcommand{\sE}{\ensuremath{\bar{E}}\xspace}

\newcommand{\tE}{\ensuremath{\hat{E}}\xspace}

\newcommand{\tS}{\ensuremath{\hat{S}}\xspace}

\newcommand{\ts}{\ensuremath{\hat{s}}\xspace}

\newcommand{\ctE}{\ensuremath{\tilde{E}}\xspace}
\newcommand{\cts}{\ensuremath{\tilde{s}}\xspace}
\newcommand{\ctS}{\ensuremath{\tilde{S}}\xspace}
\newcommand{\reals}{\mathbb{R}}

\newcommand{\absent}{\ensuremath{\sqcup}\xspace}

\newcommand{\Nat}{\ensuremath{\mathbb{N}}\xspace}

\title{Towards Verification\\ 
       of Uncertain Cyber-Physical Systems}

\author{Carna Radojicic, Christoph Grimm
\institute{TU Kaiserslautern, Germany}
\email{radojicic|grimm@cs.uni-kl.de}
\and
Axel Jantsch, Michael Rathmair
\institute{TU Wien, Austria}
\email{jantsch|rathmaier@ict.tuwien.ac.at}
}
\def\titlerunning{Towards Verification of Uncertain Cyber-Physical Systems}
\def\authorrunning{C. Radojicic, C. Grimm, A. Jantsch \& M. Rathmair}

\begin{document}
\maketitle

\begin{abstract}
Cyber-Physical Systems (CPS) pose new challenges to verification and validation that go beyond the proof of functional correctness based on high-level models. 
Particular challenges are, in particular for formal methods, its heterogeneity and scalability. 
For numerical simulation, uncertain behavior can hardly be covered in a comprehensive way which motivates the use of symbolic methods.

The paper describes an approach for symbolic simulation-based verification of CPS with uncertainties.  
We define a symbolic model and representation of uncertain computations: Affine Arithmetic Decision Diagrams.
Then we integrate this approach in the SystemC AMS simulator that supports simulation in different models of computation. 
We demonstrate the approach by analyzing a water-level monitor with uncertainties, self-diagnosis, and error-reactions. 
\end{abstract}

\section{Introduction}
\label{sec:introduction}
Cyber-Physical Systems (CPS) consist of a large number of networked, co-operating and open sub-sys\-tems~\cite{Lee08}.
This is a blessing and a curse: 
On one hand, the high number of components and their open nature make it likely that some components fail, 
are changed, or get inaccurate sensor data.
On the other hand, co-operation allows us to implement resilience that maintains dependable operation.
As CPS often implement mission-critical ecosystems and services, e.g. autonomous driving, or aviation,
it is mandatory to show that errors and deviations must not lead to a failure or to an unsafe state (fail safe). 
Even more: in CPS, the correct function shall be maintained under presence of uncertainties (fail operational).  
However, in CPS, the terms ``error'' and ``correctness'' require a new understanding.

The first reason for this  is that faults, unforeseen changes, and deviations have to be considered as likely part of normal behavior.
Therefore, we summarize all kinds of such events that cause deviations from the ideal behavior under the more general term  {\em uncertainty}. 
Due to its higher probability, the propagation and interaction of multiple uncertainties of different kinds must be verified thoroughly.
However, the complexity of  dynamic behavior, and interactions of multiple uncertainties across different domains 
are often too complex to be handled by simulation or human imagination.
It requires exhaustive methods such as symbolic simulation or model checking.

The second reason is that CPS are deployed in a more open environment than 
embedded systems. In consequence, correctness in the sense of fulfilling specified properties that
stem from design-time analysis of well-defined use-cases might be too simple.
CPS also have to cope with changing requirements or unforeseen scenarios. 
This demands for the application of adaptive methods. 

This paper pursues the following objectives: 
First, we describe how formal verification and symbolic simulation can contribute in an heterogeneous, industrial verification process for CPS. 
Second, we study the propagation and interaction of uncertainties in some models of computation (MoC) as a starting point for future work that shall provide a basis for the symbolic simulation in arbitrary combinations of different MoC. 
Implementation and examples can be downloaded from {\tt http://cps.cs.uni-kl.de/AADD}.

\subsection{Related work and contribution}
For the understanding of interacting discrete and continuous subsystems of CPS with uncertain behavior, 
research on verification of hybrid systems provides valuable insights. 
In the continuous domain, the  -- due to propagation of uncertainties modeled by non-deterministic behavior -- 
reachable state space is segmented by planes into convex geometric figures. 
Zonotopes~\cite{Girard2005,Althoff2011} and support functions~\cite{LeGuernic2009} improve scalability in non-linear systems. 
Also, affine arithmetic~\cite{Figueiredo2004} is used in particular in the verification of analog circuits and systems \cite{Grimm2005}. 
Its geometrical interpretation is similar to a zonotope, but it offers further useful properties; they in particular maintain correlation information. 
In order to yield high scalability also for discrete systems, functionally reduced AND-Inverter graphs are combined with models of linear continuous dynamics \cite{Damm2006}. 

For the discrete subsystem, this work is very similar to methods for symbolic execution of software, where control-flow introduces path conditions as discrete states. 
To cope with the path explosion problem SAT and/or SMT solvers (e.g. \cite{Saswat2008,Jeon2012}) are used to determine the reachable paths. 
Affine arithmetic has been used in this context for the static analysis of rounding errors in DSP algorithms \cite{Fang2002b} or numerical programs and even hybrid systems in \cite{Putot2013} based on splitting affine arithmetic forms (AAF) and joining them in an enclosing hull, in particular targeting stability and robustness. 

For CPS, its networked, distributed, and heterogeneous nature poses additional challenges \cite{Lee08}.
This includes in particular the use of domain-specific modeling formalisms in different parts of a CPS.
In this context, the term {\em models of computation} also became popular in the domain of modeling/simulation. 
Lee and Sangiovanni-Vincentelli~\cite{Lee2006} introduced a meta-model in which different models of computation and different means for communication and synchronization can be uniformly represented and compared.
A mostly similar, but more refined approach is proposed by Jantsch in~\cite{Jantsch2003,Jantsch2005A}.

This paper contributes two main results: 
First, we introduce an efficient yet simple method to compute with uncertain values that combines affine arithmetic with binary decision diagrams: Affine Arithmetic Decision Diagrams (AADD; Sec.~\ref{sec:computations}).
Second, we deliberately distinguish between symbolic computation with uncertain values and concrete models of computation (Sec.~\ref{sec:MoC}). 
This allows us to model CPS in a variety of models of computation without being limited to a specific one such as hybrid automata.
We demonstrate the approach by an example (Sec.~\ref{sec:example}).

\section{Verification of uncertain CPS}
\label{sec:methodology}

For software and digital systems, the underlying digital synchronous hardware platforms allow us to abstract from all physical variations in e.g. temperature, supply voltage, and to focus on the `ideal', intended behavior of algorithms.
Hence, inside this domain we can easily specify properties, prove its correctness, and trust the results. 
This does not hold once physical domains or human interactions are involved, e.g. an autonomously driving vehicle in a city. 
Therefore, for verification of CPS we have to very carefully re-think basic assumptions, methodologies, and strip down verification to  different, well-defined verification problems. 

A particular challenge for verification of CPS in this context is the presence of uncertainty. 
Uncertainty can be defined as {\em``any deviation from the unachievable ideal of completely deterministic knowledge of the relevant system''} \cite{Walker2003A}. 
Formally, uncertainty can be modeled by non-deterministic or probabilistic choice of a value from a set or range.
As uncertainties, we treat in particular unknown deviations due to modeling or implementation issues, variations in physical implementations, abstractions in models, faults, or errors. 
Furthermore, uncertainties can also come from the environment of CPS (external  uncertainties). 
External uncertainties can be operating parameters such as the temperature or humidity, but also inputs that are uncertain, e.g. jitter in clocks, but also unforeseen use case scenarios. 

\begin{figure}[htb]
\centering
\includegraphics[scale=0.65]{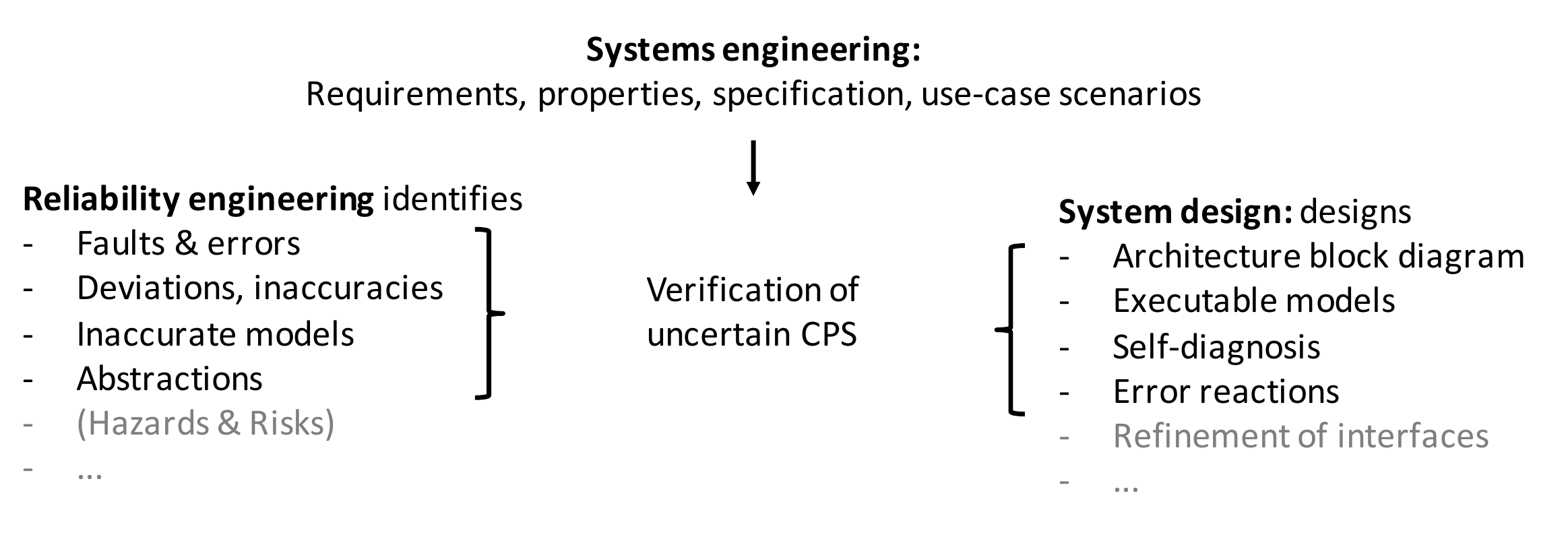}%
\caption{Overview of the development- and verification methodology.}
\label{fig:methodology}
\end{figure}

For the verification of uncertainties in CPS we assume a development process as shown in Fig.~\ref{fig:methodology}:
Systems engineering formulates, more or less formal, required properties, use case scenarios, and maybe a functional model. 
Verification objective is, at this stage, to show that the functional model fulfills the required properties and is free of inherent problems. 

The following system development consists at least of system design and reliability engineering. 
System design refines the functional model to an architecture level block diagram, and reliability engineering.
Reliability engineering identifies risks, hazards and initiating faults and deviations. 
Then it is investigated whether initiating faults or deviations can become an error or a hazard. 
If this is the case, methods for the self-diagnosis and error reactions are implemented. 
Objective of the verification of uncertain CPS is to support these tasks. 

We make the following basic assumptions for verification of uncertain CPS: 
\begin{itemize}
\item The system design is specified in a domain-specific language in the form of a block diagram. 
\item The components have parameters that model arbitrary kind of uncertainties such as the potential presence of continuous deviations or discrete faults.
\item The inputs can have parameters that model external uncertainties.
\end{itemize}
The objective of verification of uncertain CPS is to show that for arbitrary chosen uncertain values the properties of the design are within its specified ranges. 
Our approach can be considered as a bounded-time symbolic simulation with assertion checking \cite{Radojicic2013}.

Obviously, verification of uncertain CPS does not show its general correctness.
It must be complemented with other approaches that focus specifically, e.g. on concurrency issues. 
We deliberately limit our approach to one task to yield better scalability.
In the following, we first formalize the representation of uncertainties and its 
propagation in computations, and then in different models of computations. 


 

\section{Uncertainties in computations}
\label{sec:computations}
In the following, we first formalize the representation of uncertainties in symbolic computations. 
Informally, we represent uncertain values, short uncertainties, as quantities (e.g. $\tilde{x},\hat{x},\check{x}$) that depend from uncertain variables $\varepsilon_i, \x_i$. We consider these variables as the atomic, basic sources of uncertainty.  

\begin{definition}[Basic uncertainties] \sloppy
A discrete basic uncertainty is a variable $\x_i, i \in \{1, \ldots, m\} \subset \Nat$ which takes values in the set $\B=\{true, false \}$.  
Let $\X = \{\x_1, \ldots, \x_m \}$ be the set of all basic discrete uncertainties. 
A continuous basic uncertainty  is a variable $\varepsilon_j, j \in \{1, \ldots n\} \subset \Nat$ which takes values in $[-1,1]\subseteq \R$.
Let $\Epsilon  =  \{\varepsilon_1, \ldots, \varepsilon_n \}$ be the set of all basic continuous uncertainties.
$m$ and $n$ can be increased.
\end{definition}

\subsection{Propagated continuous and discrete uncertainties}

\begin{definition}[Continuous propagated uncertainty]
A continuous propagated uncertainty $\tilde{x}$ is a quantity $\tilde{x}: \Epsilon \rightarrow \R$ that describes the dependency of the real-valued result of a computation from the basic continuous uncertainties $\Epsilon$. 
\end{definition}
We represent continuous propagated uncertainties by affine arithmetic forms (AAF, \cite{Figueiredo2004}):
\begin{equation}
\tilde{x} = x_0 + \sum_{i=1}^n x_i \varepsilon_i \qquad \mbox{ with } \varepsilon_i \in [-1,1],  x_0\in \R, x_i \in \R
\label{eq:aaf}
\end{equation}
where each $x_i$ models the sensitivity of $\tilde{x}$ to the basic uncertainty $\varepsilon_i$ ($1^{st}$ order effects).
Let $\range(\tilde{x})$ of an AAF $\tilde{x}$ be an interval $[lb,ub] \subseteq \R$ with:
\begin{equation}
\range(\tilde{x})=[x_0 - \sum_{i=1}^n x_i, \quad x_0+\sum_{i=1}^n x_i]
\label{eq:aaf-range}
\end{equation}
The linear operations $(+, -, \cdot)$ where $\cdot$ corresponds to multiplication with a constant $c\in \R$ on AAF are:
\begin{equation}
c(\tilde{x} \pm \tilde{y}) = c(x_0 \pm y_0) + \sum_{i=1}^n c(x_i \pm y_i) \varepsilon_i 
\label{eq:aaf-linear}
\end{equation}
Non-linear operations $\tilde{z} = f(\tilde{x}, \tilde{y})$ are handled by a linear inclusion and  increase $n$ by 1. 
For this purpose $f(\tilde{x}, \tilde{y})$ is over-approximated by an affine form $f^a(\tilde{x},\tilde{y})=z_0 + \sum_{i=1}^n z_i \varepsilon_i$ and an additional term $z_{n+1}\varepsilon_{n+1}$: 
\begin{equation}
f(\tilde{x},\tilde{y}) \subseteq f^a(\tilde{x},\tilde{y})+z_{n+1}\varepsilon_{n+1}
\label{eq:aaf-nonlinear}
\end{equation}
The computation of $f^a(\tilde{x}, \tilde{y})$ and $z_{n+1}$ is a multi-criteria optimization problem that is solved by approximation schemes that make different trade-offs, e.g \cite{Figueiredo2004}:
\begin{itemize}
  \item {\em Minimal range approximation} -- 
    The minimal range approximation eliminates over-approximation at interval bounds (Min: range$(f(\tilde{x},\tilde{y}))$) at the cost of an increasing $z_{n+1}$. 
  \item {\em Chebychev approximation} -- The Chebychev approximation minimizes $z_{n+1}$ at the cost of increasing the range of $f^a(\tilde{x}, \tilde{y})$. 
\end{itemize}
As each non-linear operation increases $n$ by one, the number of variables grows linear with the number of nonlinear operations. 
To avoid this, we use only a single additional variable $\epsilon$ to which we add all approximation errors by assuming them as uncorrelated. 
This also guarantees safe inclusion. 

\paragraph*{Example}
As example consider two AAF $\tilde{a}=1+2\varepsilon_1, \tilde{b}=2-2\varepsilon_1+\varepsilon_2$. 
Then  $\tilde{a}+\tilde{b}=3+\varepsilon_2$, and $a \cdot b=2+2\varepsilon_1+\varepsilon_2+[-6,2]$ 
where $[-6,2]$ ensures safe inclusion; to come to an AAF with a $\varepsilon_3 x_3$ including the non-linear terms, 
we can either increase $[-6,2]$ to $[-6,6] (x_3=6)$, or increase $x_1$ and $_2$ while reducing~$x_3$.  

Note, that the geometrical interpretation of an AAF is equivalent to two zonotopes:
one including $x_{n+1}\varepsilon_{n+1}$ that gives the enclosing hull (over-approximation), and another one without $x_{n+1}\varepsilon_{n+1}$ that gives the inner hull (under-approximation). 

\begin{definition}[Discrete propagated uncertainty]
A discrete propagated uncertainty $\check{x}$ is a Boolean function $\check{x}: \X \rightarrow \B$ that describes the dependency of the ideal result of a computation from the basic discrete uncertainties $\X$. 
\end{definition}
For example, discrete propagated uncertainties can describe the possible results of a Boolean function in the potential presence of discrete faults (i.e. the $\x_i$). 
As discrete propagated uncertainties are boolean functions, we represent them by (if needed reduced) ordered binary decision diagrams (ROBDD \cite{Bryant1986}).
We assume the reader is familiar with ROBDD.
In brief, a BDD is a DAG $(V,E)$ whose terminal vertices are labeled $true$ or $false$, and whose internal vertices $v \in V$ are labeled with the variables $x_i, i \in \{ 1, \ldots, n \}$ of a boolean function $y=f(x_1, \ldots, x_n)$
and are connected with a vertex $v_t=true(v)$ and $v_f=false(v)$, depending of the value of a variable $x_i$.
If the order of its variables $x_i$ is the same on all paths, it is ordered (OBDD).
If all redundancies in form of isomorphisms and double-edges are removed, it is reduced (ROBDD). 

\subsection{Hybrid uncertainties} 

Computations consist of both discrete and continuous parts. 
This is  the case for algorithms where computations on real numbers are controlled by a discrete control flow, i.e. conditional statements and iterations. 
In the following we extend the definitions of uncertainties towards `hybrid uncertainties'.

\begin{definition}[Hybrid uncertainty]
A hybrid uncertainty $\hat{x}$ is a quantity $\hat{x}: \X  \times \Epsilon \rightarrow  \R \cup \B$ that describes the dependency of a real-valued or boolean result of a computation from the discrete and continuous basic uncertainties $\X, \Epsilon$.
\end{definition}

\begin{figure}[htb]
\centering
\includegraphics[scale=0.55]{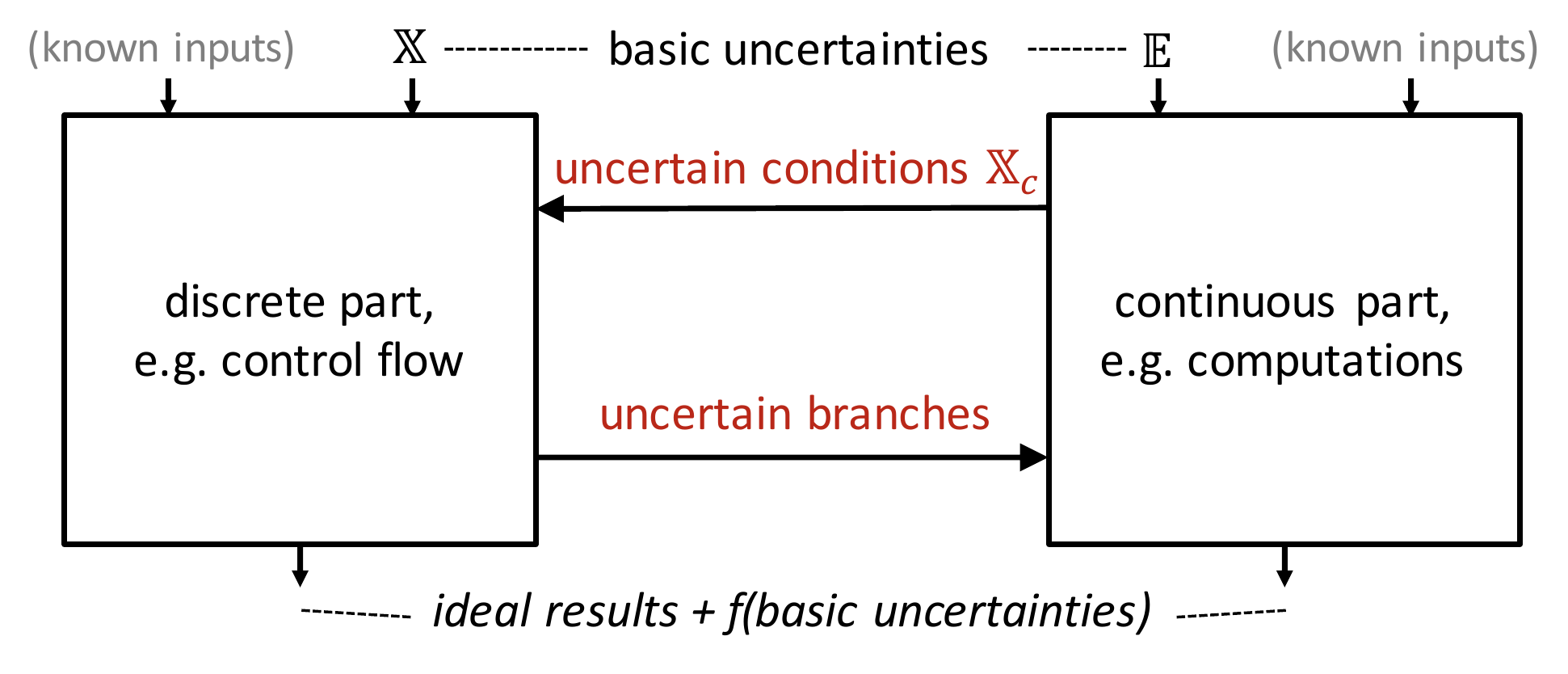}%
\caption{Interactions between discrete and continuous part in a computation.}
\label{fig:hybrid-uncertainties}
\end{figure}

We structure computations as shown by Fig.~\ref{fig:hybrid-uncertainties} into a discrete and a continuous part.
The parts interact via comparisons of continuous variables that can be uncertain (`uncertain conditions', later defined as $\X_c$), 
and via branches in the continuous part that can be uncertain.  
We consider uncertain conditions and uncertain branches as basic uncertainties for the discrete and continuous parts.

However, the uncertainty in the interaction between discrete and continuous parts must be considered when computing the  uncertainty of the overall computation (`hybrid uncertainties'). 
\cite{Putot2013} handles uncertain branches by merging all paths into a single, convex region. 
However, this introduces over-approximation that we strive to avoid. 
To come to more accurate bounds, we use the information on the interaction between the discrete and the continuous part.

\paragraph{Example} For example, consider $b=3+\epsilon_1$ and the computation {\tt if(b>3) b+=10; else b-=10;}.
Then, we cannot evaluate $b>3$ to either $true$ or $false$. 
A safe inclusion of the result would be $3+\epsilon_1+10\epsilon_2$; however, with over-approximation. 
A more accurate representation is:
$$(13+\epsilon_1 \mid (\epsilon_1>0)) \lor (-7+\epsilon_1 \mid (\epsilon_1\le 0))$$
The above representation is a Shannon expansion. 
This motivates the following representation based on ordered binary decision diagrams. 

\subsection{Affine Arithmetic Decision Diagrams}
\begin{figure}[htb]
\centering
\includegraphics[scale=0.55]{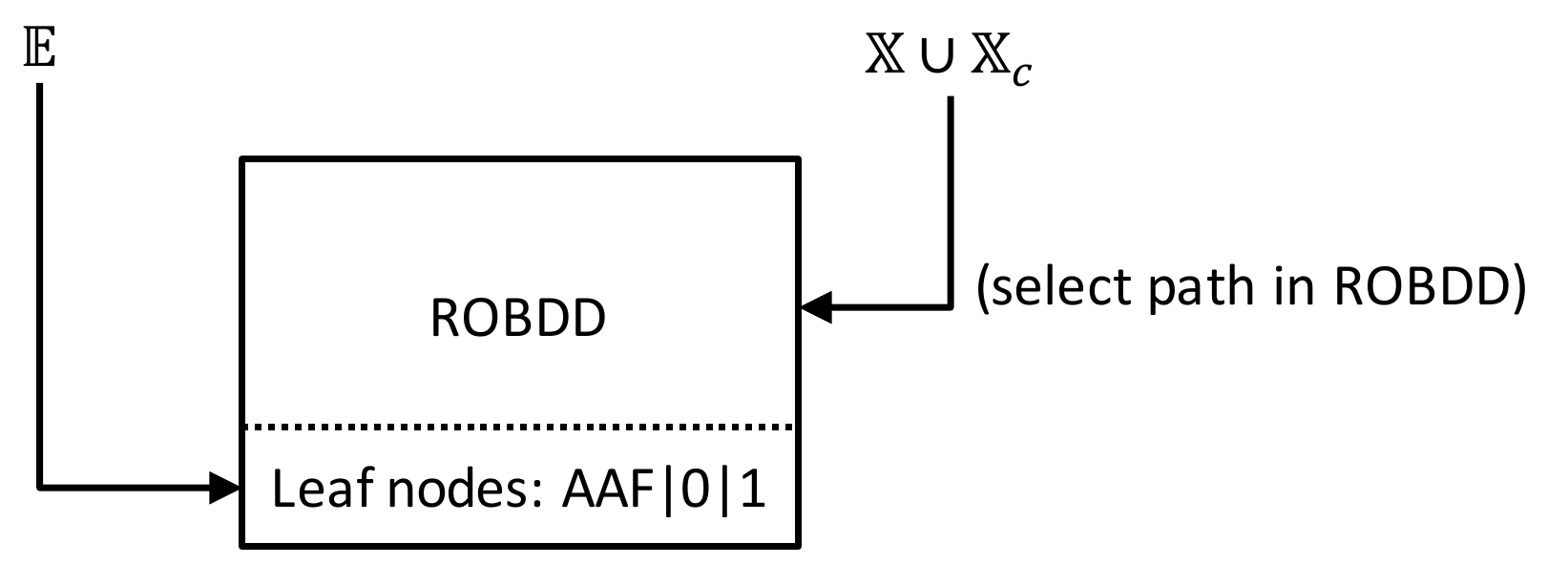}%
\caption{Overview of AADD and how they model hybrid uncertainties.}
\label{fig:aadd-overview}
\end{figure}

We represent hybrid uncertainties by affine arithmetic decision diagrams (AADD).
The idea of AADD is shown in Fig.~\ref{fig:aadd-overview}.
We use OBDD to represent the selection of an AAF by the discrete uncertainties $\X \cup \X_c$.
The OBDD can be reduced, if needed, to an ROBDD; however, we do not need it as a canonical representation. 
We label its leaf vertices with AAF in case we have an uncertain real-valued result, or with $true$ resp. $false$ if the function has a boolean result. 

\begin{definition}[Uncertain condition]\sloppy
An uncertain condition is $\x_{c}$ be a predicate whose variables are from $\Epsilon \cup \X$.
Let $\X_c$ be the set of all uncertain conditions. 
\end{definition}
For convenience, we do not distinguish between basic discrete uncertainties $\X$ and uncertain conditions $\X_c$ and abbreviate $\X\cup\X_c = \X_G = \{\x_1 \ldots \x_o\}$ (general discrete uncertainties). 

\begin{definition}[AADD]
An AADD is a directed acyclic graph $AADD=(Q,T,E_1,E_0,\X_G)$ with internal vertices $Q$, terminal vertices $T$, edges $E_1 \cup E_0$, discrete uncertainties $\X_G$, and it holds:  
\begin{itemize}
\item 
Internal vertices $v \in Q$ have two leaving edges $true(v) \in E_1$ and $false(v) \in E_0$, 
and are labeled with $index(v) = i$; each $i$ corresponds to an $\x_i \in \X_G$.
\item
AADD are ordered: $index(v_1)<index(v_2)$ for all edges $(v_1, v_2)$ from $v_1 \in Q$ to $v_2 \in Q \cup T$. 
\item 
Terminal vertices $v \in T$ are labeled with $value(v) \in \B$ for a boolean-valued hybrid uncertainty, 
or an AAF (Eq.~\ref{eq:aaf}) for a continuous-valued hybrid uncertainty. 
\end{itemize}
\end{definition}
An AADD with root $r\in Q\cup T$ represents a hybrid uncertainty $\hat{x}=f(v)$;
$f(v)$ is defined recursively: 
\begin{itemize}
\item For $v\in T$: $f(v) = value(v)$,
\item For $v\in Q$: $f(v) = \x_{index(v)} f(true(v)) \lor \overline{\x}_{index(v)} f(false(v))$ for  $\x_{index(v)}\in \X_G$
\end{itemize}
\noindent
In the following we define arithmetic, logical and relational operations on AADD.
\pagebreak

\begin{definition}[Arithmetic and binary logical operations] 
Let $\hat{x}$, $\hat{y}$ be an AADD with root vertices $v_x, v_y$.
Operations $\hat{x} \bigodot \hat{y}$ with $\bigodot: AADD \times AADD \rightarrow AADD$ are defined recursively:
\begin{enumerate}
\item 
For $v_x, v_y\in T$, the operations result in an AADD that is a terminal vertex $v$ labeled  $value(v)=value(v_x) \bigodot value(v_y)$.
For $value(v_x), value(v_y)$ of type AAF~(Eq.~\ref{eq:aaf}), $\bigodot$ is given by Eq.~\ref{eq:aaf-linear},\ref{eq:aaf-nonlinear}.
For $value(v_x), value(v_y) \in \B$, the operation $\bigodot$ are boolean functions. 
\item 
For $v_x \in T, v_y\in Q$ the result is an AADD with root $v$ and $index(v)=index(v_y)$, $false(v)=v_x \bigodot false(v_y)$ and $true(v)=v_x \bigodot true(v_y)$. 
\item 
For $v_x, v_y\in Q$ the result is an AADD with root $v$ and
\begin{itemize}
\item 
For $index(v_x)=index(v_y)$: \\
$index(v)=index(v_x)$, $false(v)=false(v_x) \bigodot false(v_y)$, $true(v)=true(v_x) \bigodot true(v_y)$.
\item 
For $index(v_x)<index(v_y)$: \\
$index(v)=index(v_x)$, $false(v)=false(v_x) \bigodot v_y$, $true(v)=true(v_x) \bigodot v_y$.
\item
For $index(v_x)>index(v_y)$: \\
$index(v)=index(v_y)$, $false(v)=v_x \bigodot false(v_y)$, $true(v)=v_x \bigodot true(v_y)$.
\end{itemize}
\end{enumerate}
\end{definition}

In the following we define relational operations. 
We define the comparison of an AADD as with $0$ as the right operand. 
Comparisons of two AADD can be transformed into this representation by subtracting the right side of the relational operator. 
\begin{definition}[Relations]
Let $\hat{x}$ be an AADD with root $v_x$. 
Then, the inequality operation $\hat{x} \oslash 0$ with $\oslash \in {<, >, ==, \leq, \geq}$ and $\oslash: AADD \times \{0\} \rightarrow AADD$, is defined recursively as:
\begin{enumerate}
\item For $v_x \in T$ the result is an AADD that consists of one vertex $v_z$. 
\begin{itemize}
\item  
For $0 \notin \range(value(v_x))$, the result $v_z$ is a terminal vertex with $value(v_z)$ defined by Table~\ref{tab:operators}. 
\item
For $0 \in \range(value(v_x))$, the result $v_z$ is an internal vertex.
$\x_{m+1}=value(v_x)\oslash 0$ is added to $\X_G$, and $index(v_z)=m+1$,
$false(v_z)=0$, and $true(v_z)=1$. 
\end{itemize}
\item For $v_x \in Q$ the result is an AADD with root $v_z$ and $index(v_z)=index(v_x)$,   $false(v_z)=false(v_x)$ and $true(v_z)=true(v_x)$.
\end{enumerate}
\end{definition}
\begin{table}[htbp]
\centering \small
\caption{Relational operators.}
\label{tab:operators}
\begin{tabular}{|l|l|}
\hline
$value(v_x) < 0$            & 
  $\begin{array}{ll}
     value(v_z)=1  &:  ub(value(v_x))<0 \\
     value(v_z)=0  &:  lb(value(v_x)) \geq 0 \\
     v_z &: otherwise
  \end{array}
 $ \\ \hline 
$value(v_x) \leq 0$            & 
  $\begin{array}{ll}
     value(v_z)=1  &:  ub(value(v_x)) \leq 0 \\
     value(v_z)=0  &:  lb(value(v_x)) > 0 \\
     v_z &: otherwise
  \end{array}
 $ \\ \hline
$value(v_x) > 0$            & 
  $\begin{array}{ll}
     value(v_z)=1  &:  lb(value(v_x)) > 0 \\
     value(v_z)=0  &:  ub(value(v_x)) \leq 0 \\
     v_z &: otherwise
  \end{array}
 $ \\ \hline
$value(v_x) \geq 0$            & 
  $\begin{array}{ll}
     value(v_z)=1  &:  lb(value(v_x)) \geq 0 \\
     value(v_z)=0  &:  ub(value(v_x)) > 0 \\
     v_z &: otherwise
  \end{array}
 $ \\ \hline
 $value(v_x) == 0$            & 
  $\begin{array}{ll}
     value(v_z)=1  &:  value(v_x) == 0 \\
     value(v_z)=0  &:  (ub(value(v_x)) < 0) \lor (lb(value(v_x))>0) \\
     v_z &: otherwise
  \end{array}
  $ \\ \hline
\end{tabular}
\end{table}
The definition of the inequality operation $\neq: AADD \times AADD \rightarrow \mathbb{R}$ is straight forward since it holds that $(\hat{x}\neq 0) =  \neg (\hat{x}==0)$.

The AAF at the terminal nodes are representing the correlations correctly, but over-approximate the range.
Accurate bounds are required for the evaluation of relational operators according to Tab.~\ref{tab:operators}. 
The over-approximation is mostly due the uncertain conditions that impose additional constraints.
To compute the exact range of $\hat{x}$ in a terminal node $v_x\in T$ with value $\tilde{x}$ we set up a system of inequations (for all indices $k$ of the nodes along the path to $v_x$): 
\begin{equation}
lb(\tilde{x}), ub(\tilde{x})
\mbox{ subject to the uncertain conditions for all internal vertexes along the path to $v_x$}  \end{equation}

In the concrete implementation of AADD, this is the LP problem that is solved using GLPK LP solver.

\begin{figure}[htb]
\begin{minipage}{0.35\textwidth} 
\paragraph*{Example}
In Fig.~\ref{fig:aadd-example} we show a computation with uncertain values {\tt AAF a(0,2), b}.
The representation by AAF is shown to the right of the C++ code. 
The statements in the conditional statement introduce conditions; 
the AADD for {\tt b} after the conditional statement is shown at the right hand side of the figure. 
\end{minipage}
\hfill
\begin{minipage}{0.62\textwidth}
    \includegraphics[scale=0.5]{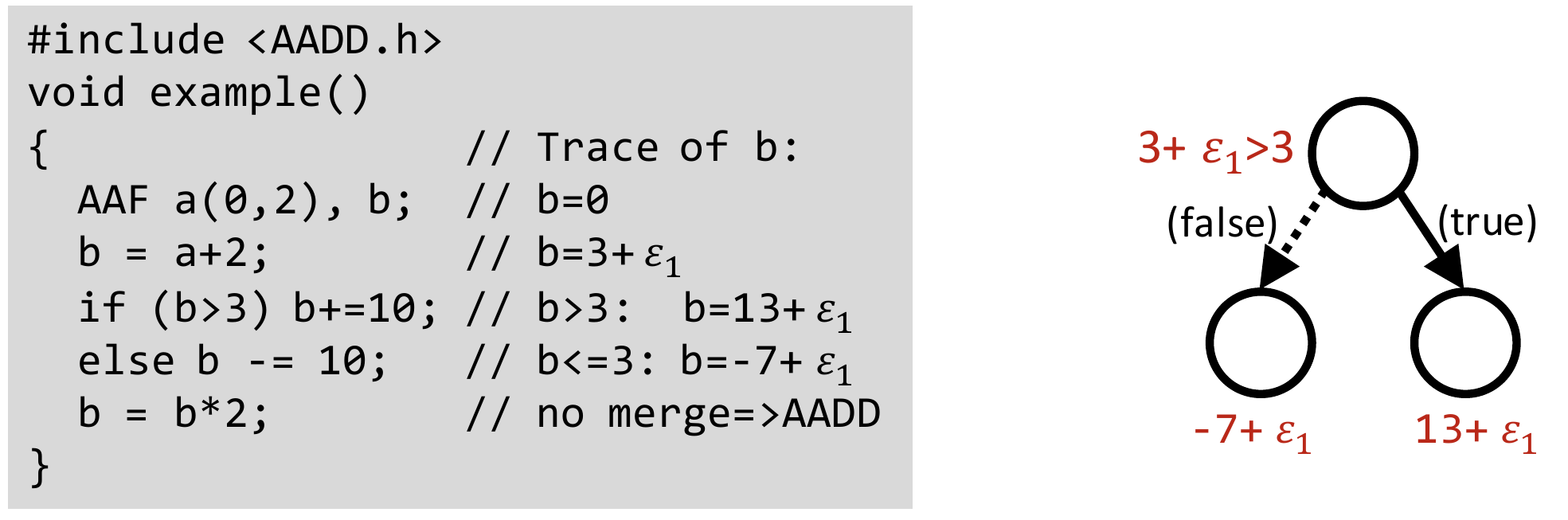}%
	\caption{Example for an AADD.}
    \label{fig:aadd-example}
\end{minipage}
\end{figure}

\section{Uncertainties in Models of Computation}
\label{sec:MoC}
So far, we have neglected communication and synchronization and focused on the pure computational kernel of processes.
In this section we study the impact of uncertainties on synchronization and process execution mechanisms that define Models of Computations (MoCs). 

\subsection{Processes and signals} \label{sec:processes-signals}

Lee and Sangiovanni-Vincentelli introduced the tagged signal model~\cite{Lee2006} as a meta-model based on which the interaction of models of computation such as dataflow, discrete and continuous timed and or synchronous languages can be modeled.
A tag is drawn from a set $T$ that models time and that may be a partially ordered set (untimed models), the integers (timed models), or the set of real numbers (continuous timed models). 
Hence, MoCs are distinguished only by the structure of signals and, to some extent, by the execution mechanics of processes.

In the following we briefly describe the model from~\cite{Jantsch2003,Jantsch2005A}  which is equivalent to the tagged signal model but represents time implicitly by the order of events rather than by explicit time stamps.
Processes communicate with each other by writing to and reading from signals.
Given is a set of values $V,$ which represents the data communicated over the signals.
$V$ may be a discrete set 
or a continuous set. 
\emph{Events} are the basic elements of signals; they consist of values and time tags. In the special case of untimed signals, the tags may be omitted and ordering of events is determined by the ordering within a signal. 
We distinguish between three different kinds of events.

\emph{Untimed events} \uE are just values without further time information beyojnd the ordering within the signal, $\uE=V$. 
\emph{Discrete timed events} \tE\   include a pseudo value $\perp$ in addition to the normal values, hence $\sE = V \cup \{\perp\}$. 
\emph{Continuous timed events} \ctE use the real numbers as time tags and thus, they can be considered as a $(v,t)$ pair with $v\in V, t\in \reals$. 

We use $E = \uE \cup \tE \cup \ctE$ and $e \in E$ to denote any kind of event.

Signals are sequences of events. 
Sequences are ordered and we use subscripts as in $e_i$ to denote the $i^{\mathrm{th}}$ event in a signal. 
E.g. a signal may be written as $\langle e_0,e_1,e_2 \rangle$. 
In general signals can be finite or infinite sequences of events and $S$ is the set of all signals. 
We also distinguish between three kinds of signals and \uS, \tS and \ctS denote the untimed, discrete timed and continuous timed signal sets, respectively, and \us, \ts and \cts designate individual untimed, discrete timed, and continuous timed signals. A particular type of signal is used in the corresponding Model of Computation, e.g. an untimed MoC contains only untimed signals.


Processes are defined as functions on signals
\[p : S \rightarrow S, \]
\noindent which means process $p$ consumes the events on its input signal and produces the events on its output signal. Since a process is a function, it is deterministic and will produce always the same output signal for a given input signal. 
They may have internal state, which means that the generated event at the output depends on the input event and the internal state of the process at that time. 
A \emph{process network} is constructed by connecting processes via their input and output signals.



\subsection{Models of Computation}

Now we are in a position to define several Models of Computation (MoC) that are popular in hardware, software or embedded systems design. 
A MoC determines execution mechanics that activates a process based on specific conditions, e.g. availability of inputs. 

\begin{definition}[Untimed MoC]
An untimed MoC is the set of all process networks where all processes communicate with each other with untimed signals $\us\in \uS$ only.
\end{definition}
Processes can be executed once all inputs are available. 
Outputs can be written depending on values computed by processes.
An example is the Kahn Process Networks (KPN) MoC \cite{Kahn1974}.

\begin{definition}[Static Dataflow] \label{def:static}
Static Dataflow is an untimed MoC where each process consumes and produces always the same number of events in each evaluation cycle.
\end{definition}
Different processes may consume and produce different number of events. 
In  the literature this static dataflow is often also called synchronous dataflow.
Process execution is done in a static schedule that can be computed before execution.


\begin{definition}[Discrete Time MoC]
Discrete Time MoC is the set of all process networks where all processes communicate with each other with timed signals $\ts\in \tS$.
\end{definition}
Processes can be activated by arbitrary conditions of time, inputs, or internal states.
Examples for discrete time MoC are the discrete-event simulation semantics of VHDL or SystemC. 

\begin{definition}[Continuous Time MoC] \label{def:ct}
Continuous time MoC describes processes by differential and algebraic equations (DAE). 
Signals are continuous timed signals $\cts\in \ctS$ that represent the solutions of the DAEs. 
\end{definition}
In the continuous-time MoC, the process execution is controlled by a `solver'. 
The solver can select discrete time steps in order to approximate the ideal, continuous time signals while minimizing the error due to discretization.
Values of events for arbitrary times can be computed by interpolation if needed.

Note, that the set $V$ of values transported by signals and processed by processes is irrelevant for the definition of MoCs. 
$V$ may be an arbitrary set, binary, discrete, continuous, structured, etc.
This obliviousness allows us to extend these MoCs to values with uncertainties. 
However, uncertain values might have impact on the concrete behavior of models in different MoCs.
 
\subsection{Uncertain values and their impact on MoCs}

In the case of uncertain values, the set $V$ becomes a set of continuous, discontinuous, or hybrid (propagated) uncertainties.
Then, the computation of the function $p$ that defines the behavior of processes can be computed and represented as described in Sec.~\ref{sec:computations}. 
However, as conditions on values can trigger the execution of processes, we have to study the impact of uncertain values. 
For this purpose we distinguish static and dynamic MoC activation: 
\begin{definition}[Static/Dynamic MoC] 
MoC are called static if the order of all events and their times are independent from the values of the computations. 
MoCs that are not static are dynamic. 
\end{definition}

Static dataflow is a static MoC:  
by definition~\ref{def:static}, a constant number of events is produced and consumed at each process activation. 

The continuous time MoC can also be considered as a static MoC.
In continuous time MoC, there are events for every time drawn from $\reals$; we consider the way how they are computed as an implementation detail. 

Untimed MoC and discrete timed MoC are dynamic MoC, if they are not further restricted.
As proof we give one example of a process (in pseudo-code) that shows it is not static: 
\begin{center}
\begin{minipage}{12cm}
\begin{lstlisting}[lineskip=-8pt]
  if(condition_on_value) // can be uncertain
    write(signal_event_that_activates_process); 
  else 
    do_nothing;
\end{lstlisting}
\end{minipage}
\end{center}

Nevertheless, many models and implementations in these MoC can be considered as static if they avoid activation of processes depending on conditions on values. 

If we have static MoCs or models that can fulfill the requirements for a static MoC, we can do symbolic simulation by replacing the set of values $V$ with symbolic representations such as AADD. 
Static MoCs (or static models) pave the path for a very easy implementation in design and modeling languages that support abstract data types and operator overloading; examples for such languages are C++, VHDL, SystemC. 
Then, we can replace pre-defined data types with an AADD-based data type and leave the symbolic simulation to overloaded operators. 

For dynamic MoC in the general case we have to consider that process execution can become uncertain.
This leads to a number of problems that we have not yet completely solved: 
\begin{itemize}
\item First, a process, at the same time, must be executed and not executed. 
We can solve this by adding a symbol $\absent$ to the set of reachable values represented by an AADD.
Then, a process, when executed can compute outputs also for this case (that is, doing nothing) in addition to the other symbolic manipulations on AADD; however this is seen as future work.  
\item A fundamental issue is the interface between continuous-time MoC and other MoC that leads to a non-computable problem in general. 
      Here, uncertainty of values in conditions leads to an uncertain time of process activation; 
      to cope with this issue is rather a modeling challenge and can be avoided in many cases.
\end{itemize}

\subsection{Turning the SystemC AMS simulator into a symbolic simulator}
SystemC AMS (\cite{Barnasconi2013a,Vachoux2005}, IEEE Std 1666.1-2016) targets the modeling of highly complex mixed discrete/continuous systems with a focus on a high simulation performance. 
It extends SystemC's timed (discrete-event) MoC  with support for the static dataflow MoC, and a continuous time MoC. 
By the interaction of the untimed static dataflow with the timed signals of the other MoCs its samples get timed semantics; therefore, in \cite{Barnasconi2013a,Vachoux2005} we used the term {\em timed dataflow (TDF)}. 
SystemC AMS uses TDF as a coordinating MoC that controls the interactions between different MoC and executes them in constant time steps.  
Its processes can be specified by 
\begin{itemize}
\item C++ code that models e.g. computations of embedded software. 
\item Transfer functions, linear differential equations, or static nonlinear functions.
\item Linear  circuits and switches that model sensors and actuators.
\end{itemize}
To enable industrial application, SystemC AMS provides a framework that that yields predictable behavior and scalability. 
\footnote{The standardization was clearly industry-driven with focus on scalability. Hybrid automata in contrast, with good reason, focus on the fundamental issues with a penalty in terms of scalability.}
Predictable behavior means that the behavior of models is intuitive; 
this is achieved by using TDF as a coordinating MoC in hierarchical models.
For the sake of scalability, SystemC AMS by default:
\begin{itemize}
\item Uses MoC that are scalable and numerically robust: data-flow, and linear continuous-time models.   
\item Does not compute threshold-crossing of continuous quantities; continuous quantities are sampled. 
\end{itemize}
However, SystemC AMS gives users the choice where, and how to solve problems that are known to lack scalability. 
For this purpose, SystemC AMS provides the user with interfaces to allow them to select time steps in a dynamic way, to compute thresholds accurately, and to add solvers for non-linear differential equations if needed. 

\begin{figure}[htb]
\centering
\includegraphics[width=4.9 in]{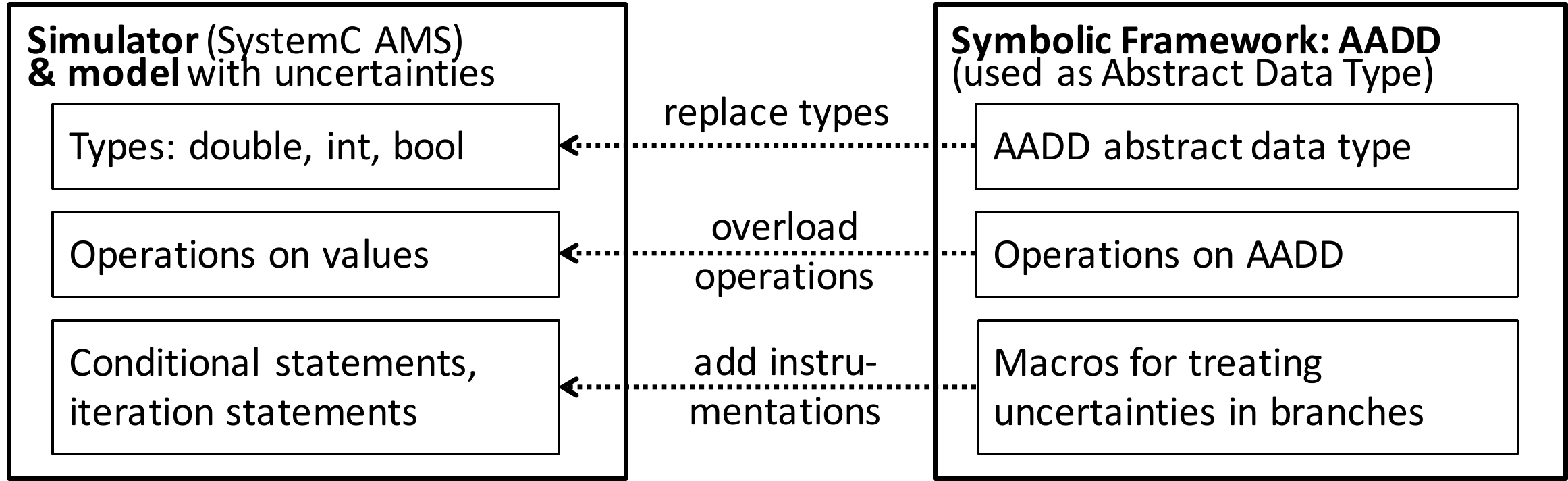}
\caption{Overview of implementation based on the SystemC AMS reference simulator.\label{fig:implementation}}
\end{figure}

We extended Coseda's implementation of the numeric SystemC AMS simulator \cite{SystemCAMSimplementation} to a symbolic simulator.
The underlying MoC are static MoC provided the C++ code does not fork additional processes or uses shared variables. 
Like the SystemC AMS standard we give the user the choice where to use symbolic simulation. 
For this purpose, we provide an abstract data type {\tt AADD} that provides AADD, overloaded operations on it, and some macros that support the interactive choice between symbolic and numeric execution for branches in loops and conditional statements.
Figure~\ref{fig:implementation} gives an overview.

\subsection{Scalability: is symbolic simulation with AADD worth the effort?}
The use of AADD imposes some overhead. 
In the following we discuss the overhead to a single, numerical simulation run and the fidelity of the results, and compare complexity with multi-run simulation based approaches that strive to yield similar results (Table~\ref{tab:scale1}).

\begin{table}[htbp]
\center
\caption{Scalability and fidelity of numerical vs. AADD-based symbolic simulation.}
\small
\begin{tabular}{|l|l|l|l|l|} \hline
                    & \multicolumn{2}{|c|}{Multi-run simulation, continuous} & \multicolumn{2}{|c|}{Symbolic with AADD} \\                 
                    & \bf MC analysis          &  \bf  WC/EVA analysis       & \bf continuous          &  \bf discrete    \\ \hline 
  Scala-            &  quadratic(confidence),  &  Worst case: $>$ exp.       & linear (\#cont. uncert.) &  exp. (\#disc. uncert.)     \\ 
  bility            &  const(\#cont. uncert.)  &  Typical case: linear       &                         &                     \\ \hline
  Fidelity          &  under-approximation     &  under-approximation        & over-approximation      &    over-approximation   \\ 
\hline
\end{tabular}
\label{tab:scale1}
\end{table}

For  numerical simulation-based approaches of continuous systems, Table~\ref{tab:scale1} gives the number of numerical simulation runs that are needed to achieve comparable, yet not comprehensive results by semi-formal approaches.
We compare Monte Carlo analysis (MC),  or Worst-Case (WC) analysis by Extreme Value Analysis (EVA). 
MC analysis has the advantage that the number of simulation runs does not grow with the number of uncertainties; it grows quadratically with the desired confidence. 
WC/EVA analysis tries to find worst-case corners of properties by checking possible combinations of corners of uncertainties. 
Various heuristics find under-approximations more efficiently in particular for linear systems, but it is impossible to get comprehensive results in the general case.
For mixed discrete/continuous systems resp. uncertain CPS, we have to visit each reachable discrete mode and do MC, WC or EVA analysis in it for comprehensive results. 

For AADD-based symbolic simulation the overhead depends on the number of uncertainties.
In the continuous domain, AADD-based symbolic simulation introduces a constant overhead that only grows with the number of uncertainties; however, over-approxima\-tion can be an issue for non-linear systems. 
In the discrete domain, we in general have the problem that there is an exponential growth of possible execution paths (path explosion problem). 
For AADD, the size (and the run-time) grows with the size of the (R)OBDD.
It {\em can} grow exponentially in the worst case with the number of basic discrete uncertainties; 
however, it does not in many cases which is a well-known issue in the research on ROBDD.

A problem is that reduction of OBDD to ROBDD costs a lot of time.
Therefore, we use OBDD or even a simple array structure as implementation in~\cite{Radojicic2016} that we call XAAF.
This is still efficient for some thousand reached discrete states (leaf nodes of AAF). 
Scalability using the array-based implementation is shown by examples of a $\Sigma\Delta$-converter in~\cite{Radojicic2016} and a PLL in~\cite{Barke2016}.  
\section{Analysis of an uncertain water-level monitor with error-reactions}
\label{sec:example}

\subsection{Modeling the water level monitor in SystemC AMS}
We have characterized cyber-physical systems in Sec.~\ref{sec:introduction} by its open and networked nature that, due to the increased likelihood of deviations or faults, demands for means to compensate such uncertainties. 
For demonstration, we extend the water level monitor model from ~\cite{Alur1995} in that direction. 
The water level monitor consists of a tank of which the water level falls with a rate of $falling=2$ in./sec.
A pump can, if switched on, add an incoming flow so that in sum the water level increases with a rate of $rising=1$ in./sec.
Two sensors indicate whether a level of 5 in. (empty) or 10 in. (full) is reached.
The water level must not fall below 1 in. or rise above 12 in. 

\begin{figure}[htb]
\centerline{\includegraphics[width=0.55 \textwidth]{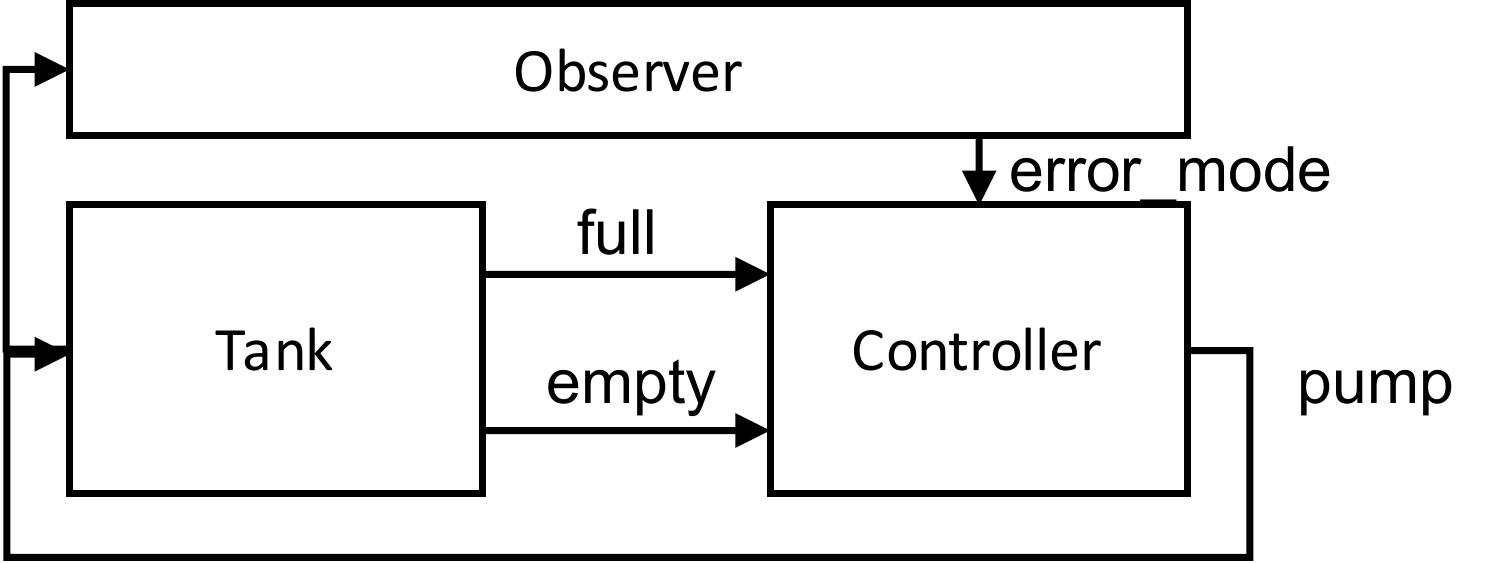}}
\caption{The cyber-physical water level monitor with observer and a controller with fail-safe mode.}
\label{fig:waterlevelmonitor}
\end{figure}
Compared with the water level monitor from ~\cite{Alur1995}, we added in Fig.~\ref{fig:waterlevelmonitor} the following uncertainties: 
\begin{itemize}
\item  Continuous uncertainties in the rates of the water tank: $falling=2+0.1\epsilon_1$, $rising=1+0.1\epsilon_2$.
\item  A discrete uncertainty: the possibility that the sensor in the tank does not indicate, when 10 in. are reached as an initiating fault: $full=\x_1$. 
\end{itemize}
Obviously, one can find and add more uncertainties. 
Furthermore, there is a process (Observer), written in C++, that shall detect the discrete fault and signal the controller to go into a fail-safe error mode. 
It is invoked every second and checks whether the pump is on for more than 10 sec. (self-diagnosis) and then signals via $error\_mode$ that the controller shall go into a fail-safe state:  
\begin{center}
\begin{minipage}{12cm}
\begin{lstlisting}[lineskip={-8.0pt}]
  void processing()     // Activated every 1.0 sec.
  {
    if (error_mode == false)
    {
      if (pump==true) {
        timer += 1;     // We take the time ... 
        if (timer > 10) error_mode = true;
      }
      else timer = 0; 
    }
  }
\end{lstlisting}
\end{minipage}
\end{center}
In the fail-safe state, the controller  limits the time the pump is on. 
We specified the water level model using the TDF MoC of SystemC AMS. 
The controller process (also written in C++) is activated every 0.1 sec. and samples the water level sensors. 
The observer process is activated every 1.0 sec. 

For symbolic simulation we have to modify the SystemC AMS model, supported by macro definitions.
This affects in particular  branches in control flow. 
Here, we have to consider that the result of a branch condition can be $true$, $false$, but as well uncertain. 
In the first two cases, the code remains unchanged. 
The third case occurs for comparisons that depend on at least one AADD. 
Then have to apply a code instrumentation that adds an uncertain condition;  
details on the code instrumentations are described in ~\cite{Radojicic2016}. 

\subsection{Symbolic simulation of the uncertain water level monitor}
Objective of the verification of uncertain CPS is to show that in the presence of the uncertainties (inaccuracies, initiating faults), self-diagnosis and error reaction ensure that no unsafe state (here: a level above 15 in.) is reached. 
For this purpose we used symbolic simulation bounded to 40 sec. time with assertion checking. 

\begin{figure}[t!]
\centering
\subfloat{\includegraphics[width=3 in]{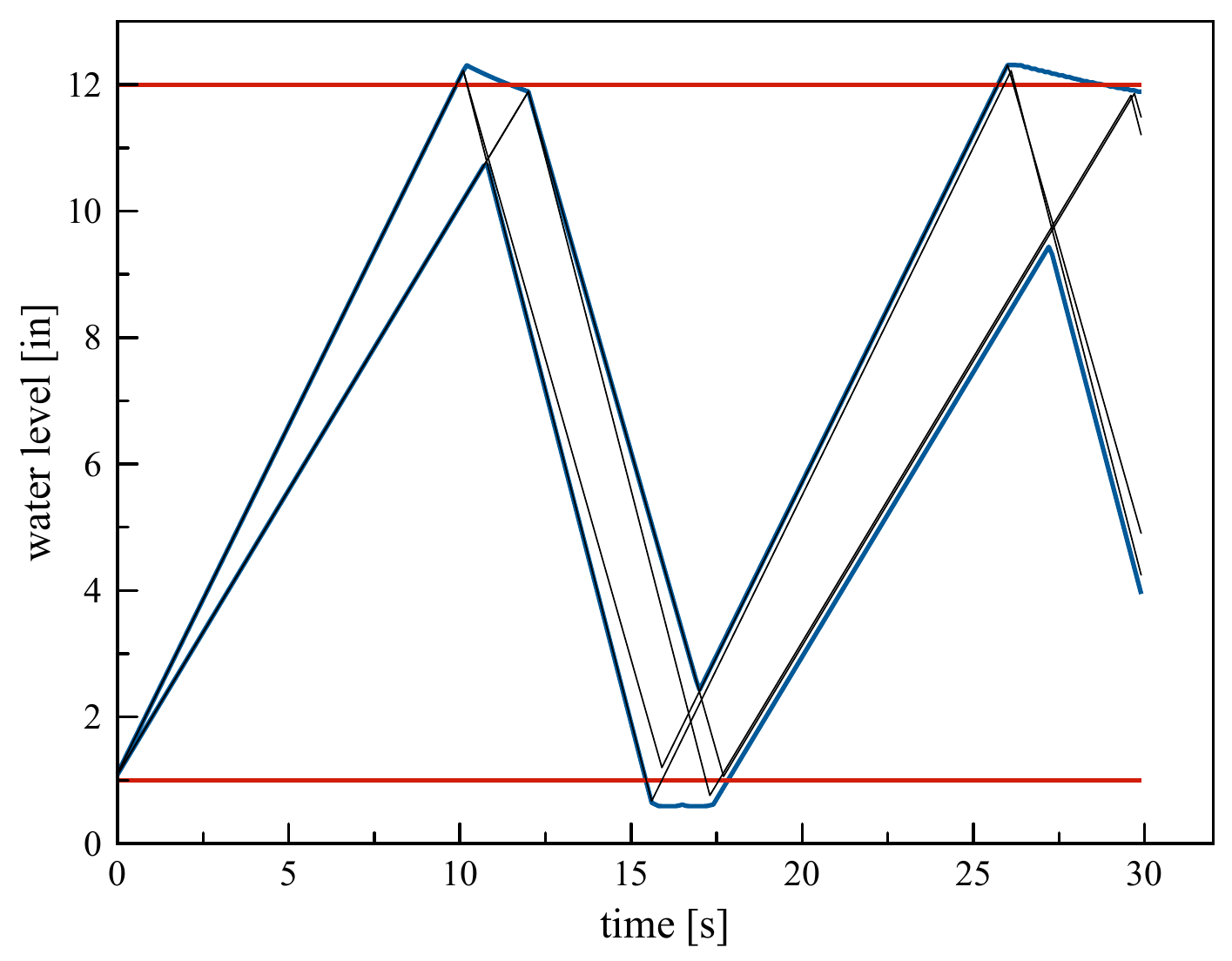}%
\label{fig:no-error}}
\hfill
\subfloat{\includegraphics[width=3 in]{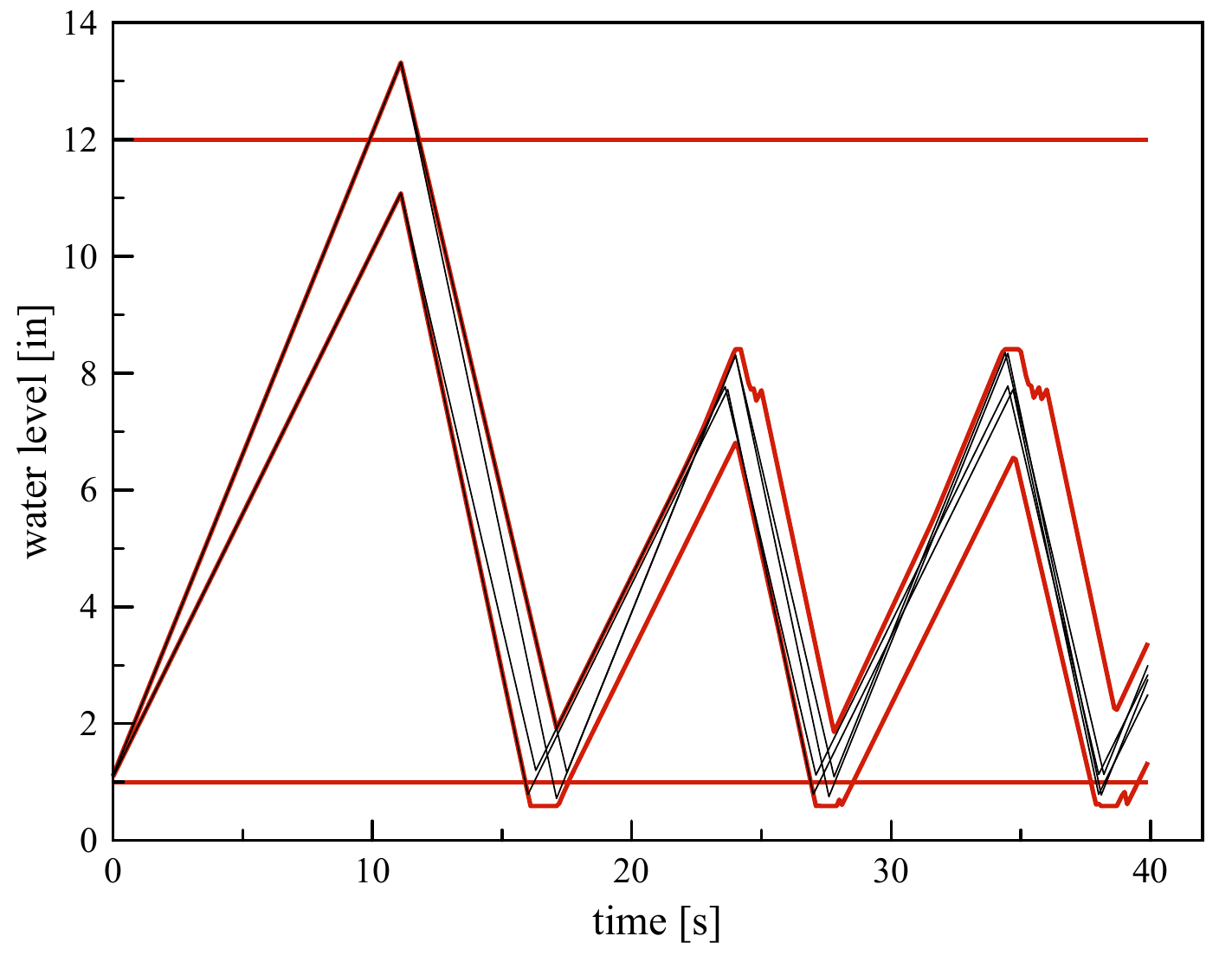}%
\label{fig:with-error}}
\caption{Symbolic simulation results (worst case borders plotted) compared with numeric worst case (by extreme value analysis) analysis results.}
\label{fig:symbolicsimulation}
\end{figure}

In figure~\ref{fig:symbolicsimulation} we show the results of symbolic simulation of the water level monitor for the cases with and without the discrete error. 
The total time required for one symbolic run of the water level monitor without observer is 6.8 sec.; it resulted in an AADD with 387 leaf nodes.
The result is shown in Fig.~\ref{fig:no-error}, left. 
As it can be seen, the inaccuracies in the rates can lead to an error. 
The corner-case simulation results (that my be an under-approximation) are plotted black, the range of the water level is plotted in blue.  
For the water-level monitor and the observer, with injecting a deterministic error in a sensor, we get a run-time of 30 sec. and an AADD with 945 leaf nodes: once the error is detected, the controller goes in a state in which he limits the on-time of the pump (Fig.~\ref{fig:with-error}, right only shows this case). 
Note, that due the interaction between water level monitor, and all of its uncertainties is complex; i.e. it must be ensured that in any case the error mode is triggered by any corner case where there is no fault (which is not the case for the given paramters); symbolic simulation assists in parameterization, design, and debugging.

\subsection{Discussion}
The example above does not show scalability to large problem sizes; this was not the intention. 
It shows that an how the approach contributes to scalability towards applicability: first,  we can (re-)use existing models or C-code.  
Second,  it shows that it is useful to solve the problem of analyzing the robustness of CPS under the presence of uncertainties including discrete faults.
This is a domain, where pure numerical simulation fails as it needs an exponential number of simulation runs. 

Lessons learned from the concrete implementation shows that run-times of symbolic simulation runs very much on details of models; e.g. we used timers that were not necessary (not wrong, but simply not needed) which doubled the run-time. 
Leaving modeling issues aside, the operations on AADD are implemented in an inefficient way and clearly need further research, while the LP problem we identified significantly contributes to keep over-approximation very little.
Interesting would in that context  be the use of a SAT/SMT solver to represent the functional dependabilities while maintaining the LP problem and the solver at the interface between discrete and continuous components.

\section{Conclusion}

We have described the problem of verification of uncertain CPS. 
As two main results we have introduced first AADD as a model that allows us compute with uncertainties, independent from a particular model of computation. 
Second, we have shown how this allows us to implement a symbolic simulator that is not strictly bound to a particular model of computation. 
The following aspects contribute to scalability towards larger problem sizes and, in particular, application in industrial development flows: 
\begin{enumerate}

\item 
The approach to separate computation with uncertainties from simulation using signals and processes in different models of computation allows us to bring symbolic simulation easily into pre-existing frameworks. 
The re-use of existing models, simulators, and other verification infrastructure from numerical simulation becomes easier, at least for static models of computation. 
Currently we support SystemC and its AMS extensions; in student's work also Labview has been enabled for symbolic simulation.  

\item 
The flexibility gained by not being bound to a particular model of computation (e.g. hybrid automata) makes it easier to formulate models that avoid known issues for scalability.
In the model of the example, we could hence make the following abstractions: 
\begin{itemize}
\item The implementation of the controller and the sensors is discrete; we have hence chosen an interface between continuous and discrete parts that uses sampling -- this avoids the need to determine threshold crossing. 
\item We used the static data flow model of computation (together with continuous-time) for the overall structure  of the model.
\end{itemize}

\item AADD combine affine arithmetic with (R)OBDD. 
Both are know to be efficient representations in the discrete and continuous domain. 
In particular the proposed use of an LP solver as described in Sec.~\ref{sec:computations} significantly reduces over-approximation even for larger numbers (i.e. 1000s) of leaf nodes (reached states). 
To foster higher scalability, operations on AADD must be further optimized to proceed to complex software systems.
While (R)OBDD and affine arithmetic are each maybe not the most efficient representations, they are at least common and well-understood, and vast research is available to increase efficiency. 
\end{enumerate}

\subsection{Future work}

Currently, we limit the implementation of the symbolic simulation to static MoCs.
This is useful as it allows us the integration of AADD in existing simulators without changing their execution semantics. 
However, this limitation is too strict if one would like to check whether deadlocks or race conditions are due to uncertainties. 
A first idea to  allow symbolic simulation also of dynamic MoCs would be to integrate a symbol for $\absent$ for that case; operators on AADD could then do nothing in that case. 

We currently only support uncertainties that are modeled by non-deterministic choice of a set or a range. 
However, in particular when modeling many possible discrete errors one is also interested in the probability of a resulting hazard. 
Ongoing work targets the extension towards probabilistic uncertainties. 
We have described a first approach in that direction in~\cite{Grimm2017}.
Also, Olbrich's distribution arithmetic~\cite{Olbrich2008} would be a useful starting point in that context.

\subsubsection*{Acknowlegement} \small
This work is funded, in part, within the ANCONA project (16ES021) within the 
program IKT 2020 by the German Ministry of Education and Research (BMBF) and by 
Robert Bosch AG, Intel AG, and Mentor Graphics GmbH. 
\vfill

\pagebreak

\bibliographystyle{eptcs}
\bibliography{cps,aaf,misc,resilience,powerprofiling}
\end{document}